\documentclass[12pt]{article}

\usepackage[english]{babel}											
\usepackage{amsmath}
\usepackage{amssymb}
\usepackage{txfonts}
\usepackage{hyperref} 
\usepackage{graphicx}

\providecommand{\keywords}[1]{\textbf{\textit{Keywords---}} #1}

\begin{document}
{\renewcommand{\thefootnote}{\fnsymbol{footnote}}
\begin{center}
{\LARGE Exploiting Anyonic Behavior of Quasicrystals \\ for Topological Quantum Computing}\\
\vspace{1.5em}
Marcelo Amaral,\footnote{Correspondence: {\tt Marcelo@QuantumGravityResearch.org}}
David Chester,
Fang Fang,
and Klee Irwin
\\
\vspace{0.5em}
Quantum Gravity Research,\\
Los Angeles, CA 90290, USA\\
\vspace{1.5em}
\end{center}
}

\setcounter{footnote}{0}

\begin{abstract}
We show that quasicrystals exhibit anyonic behavior that can be used for topological quantum computing. In particular, we study a correspondence between the fusion Hilbert spaces of the simplest non-abelian anyon, the Fibonacci anyons, and the tiling spaces of a class of quasicrystals, which includes the one dimensional Fibonacci chain and the two dimensional Penrose tiling. A possible encoding on tiling spaces of topological quantum information processing is also discussed.
\end{abstract}



\section{Introduction}
\label{intro}

While quantum computers have been experimentally realized, obtaining large-scale quantum computers still remains a challenge. Since qubits are very sensitive to the environment, it is necessary to solve the problem of decoherence \cite{Nielsen2011}. Software algorithms have been proposed by major players in the field; see citations within Ref.~\cite{Djordjevic2021}. A different seminal solution is to use hardware-level error correction via topological quantum computation (TQC) \cite{Pachos,Wang}. In particular, non-abelian anyons can provide universal quantum computation \cite{Pachos}. Theoretically, low dimensional anyonic systems are a hallmark topological phase of matter, which could be used for TQC if a concrete implementation could be achieved. While abelian anyons have been experimentally realized \cite{Bartolomei2020}, concrete evidence of non-abelian anyons still remains elusive.

Chern-Simons theory applied to the fractional quantum Hall effect and lattice models such as the toric code are theoretical frameworks for using anyons for TQC \cite{Pachos,Wang}. These systems support emergent quasiparticle excitations that show anyonic or fractional statistics.  The fusion rules and braid properties of anyons are useful for implementing TQC. The quasiparticles that encode the topological information define the structure of the fusion Hilbert space. In the Chern-Simons theory, anyons are classified by a integer parameter called the level $k$, which appears in the action of the theory. There are an infinity of levels; $k=2$ defines Abelian anyons, while greater levels define non-Abelian anyons. 
The Fibonacci anyon is the quintessential and simplest non-abelian anyon at level $k=3$ \cite{Pachos,Wang}. For our purposes, the fusion Hilbert space for Fibonacci anyons is described by the Fibonacci $C^*$-algebra  \cite{Marcolli2015}.

Due to the potential of TQC and the experimental difficulty of implementing non-Abelian anyons, it is worth understanding what forms of TQC are possible in general.
Previously, we co-authored a non-anyonic proposal of TQC from three-dimensional topology \cite{MPQGR1} and discussed their associated character varieties \cite{Planat2022}.
Here, we study quasicrystals described by the geometric cut-and-project method \cite{BaakeGrimm}. The aim is to show that tiling spaces associated with quasicrystals exhibit anyonic behavior, which can lead to TQC implementations. More specifically, we aim to establish lower dimensional quasicrystals as a new candidate to implement TQC.

Although crystallographic materials have well-developed theories, mainly Bloch and Floquet's theory, these theories do not work properly for topological aspects of quasicrystals due to the lack of translational symmetry \cite{Bellissard1992}. Nevertheless, the connection between lower dimensional quasicrystals with higher dimensional lattices allows us to adapt and to use, in a limited way, the known crystallographic theories considering subspaces of the higher dimensional Hilbert spaces. The physics of aperiodic order is a growing field of research  \cite{Bellissard1992,Kohmoto1986,Sutherland1986,Fujiwara1989,Luck1989,suto1989,benza1991,kaliteevski2000,Florescu2009,Kalugin2014,Tanese2014,Gambaudo2014,Mace2017,Mace2019,Sen2020,Baggioli2020,Jagannathan2021}. Topological superconductors have been investigated in quasicrystals, suggesting that they can exhibit topological phases of matter \cite{Indubala2013,Ghadimi2017,Varjas2019,Cao2020,Duncan2020,Liu2021,Hua2021,Fraxanet2021,Rosa2021,Sarangi2021,Fan2022}.

We present a connection between anyons and one and two dimensional quasicrystals, such as the 5-fold Penrose tiling, by the isomorphism between the anyonic fusion Hilbert space and the subspaces of lattices Hilbert spaces describing quasicrystal tiling spaces. Both spaces have dimensions growing with the Fibonacci sequence. A theorem from functional analysis says that two Hilbert spaces are isomorphic if and only if they have the same dimension. We propose that these subspaces are fusion Hilbert spaces and show an isomorphism between the Fibonacci $C^*$-algebra of Fibonacci anyons and a $C^*$-algebra associated with the tiling spaces of quasicrystals. 
The $C^*$-algebra of interest allows for the implementation of representations of the braid group necessary for topological quantum computing.

This paper is organized as follows: in Section \ref{sec:1}  we review and discuss elements of anyonic fusion Hilbert spaces and the Fibonacci  $C^*$-algebras to establish the correspondence with tiling spaces of quasicrystals. In Section \ref{sec:2} we discuss information processing aspects of tiling spaces. We present discussions and implications in Section \ref{disc}.

\section{Correspondence Between Anyons and Quasicrystals}
\label{sec:1}

The quintessential and simplest non-abelian anyon is the Fibonacci anyon \cite{Pachos,Wang}. We will show the isomorphism between anyonic fusion Hilbert spaces and quasicrystalline Hilbert spaces at the level of the Fibonacci anyons and Fibonacci quasicrystals, namely the one-dimensional Fibonacci chain and the five-fold two-dimensional Penrose tiling. The name Fibonacci in Fibonacci anyons is due the dimensions of their Hilbert spaces being a well-known Fibonacci number and in the case of the mentioned quasicrystals, we will show that they have the same behavior justifying the name Fibonacci.

\subsection{Fibonacci Anyons and Fibonacci $C^*$-algebra}
\label{subsec:11}

There are different ways to describe anyons, including the Chern-Simons (CS) theory and lattice Hamiltonian approach \cite{Pachos,Wang}. In the case of CS theory, it is well known that in (2+1) dimensions there is an additional gauge invariant term that can be added to the Maxwell or Yang-Mills Lagrangian. This CS term is topological, as it doesn't depend on the metric \cite{Pachos,Elitzur:1989nr}. At low temperatures, this term dominates. In the non-abelian case, the action is invariant under $SU(2) \cong Spin(3)$ and can be written as a Gauss constraint on a wave functional of the gauge fields. 

In the presence of sources (representations of a Lie algebra), anyonic behavior, such as fusion and braid dynamics, can be found with sufficient control of the low-temperature Hamiltonian, Lagrangian, or Gauss constraints. The degenerate ground state of the effective theory associated to the CS sources form the so-called fusion Hilbert space, which is proposed as a fault tolerant topological quantum computing substrate. In the case of Fibonacci anyons, the sources can be only in the two lower dimensional representations of $SO(3)$, the spin-0 and spin-1 representations, with fusion rules 
\begin{align}
1\otimes1 & =0\oplus1\nonumber \\
0\otimes1 & =1\nonumber \\
1\otimes0 & =1.\label{eqfibanyonfusionrules}
\end{align}
If we have $N$ spin-1 sources representations and start to fuse them, they can build different fusion paths that can lead to either spin-1 or spin-0 representations with certain probabilities. The different paths to fuse the $N$ spin-1 sources to only one spin-1 or spin-0 can be seen as states in a fusion Hilbert space $H_N$, where its dimension grows with the number of original spin-1 sources and it is given by the Fibonacci sequence, ($(0,1,)1,2,3,5,8,13,...,Fib(N+1)$) \cite{Trebst2008}, i.e., $H_N=\mathbb{C}^{Fib(N+1)}$, where $Fib(N+1)$ is the $N+1$th Fibonacci number. 

Rotating one physical source around the other is equivalent to an operation in the fusion Hilbert space described by the so-called braid operators (higher-dimensional representations of the braid group), which leads to non-trivial statistics given the necessary quantum evolution for topological quantum computation. Explicit construction of braid operators, $B$, are given as examples in \cite[Sections 2.4 and 2.5]{Trebst2008} through the so-called $F$-matrices and $R$-matrices operating in the fusion Hilbert space. For the case of fusing two anyons into a third one, this process is a five-dimensional space and the matrices explicitly in a suitable base can be given
\begin{align}
R & =diag(e^{4\pi i/5},e^{-3\pi i/5},e^{-3\pi i/5},e^{4\pi i/5},e^{-3\pi i/5}),\nonumber \\
F & =\left[\begin{array}{ccccc}
1\\
 & 1\\
 &  & 1\\
 &  &  & \phi^{-1} & \phi^{-1/2}\\
 &  &  & \phi^{-1/2} & -\phi^{-1}
\end{array}\right]\label{eq:RFmatrices}
\end{align}
with $B=FRF^{-1}$ and $\phi=2\cos(\frac{\pi}{5})\thickapprox1.618$, the golden ratio.

More details on Fibonacci anyons are well known and can be found in Ref.~\cite{Trebst2008} and references therein. Less known is the isomorphism of the fusion Hilbert spaces with representations of certain $C^*$-algebras, in particular the so-called Fibonacci $C^*$-algebra \cite{Marcolli2015}.
In \cite{Marcolli2015}, it is shown that the fusion rules determine the data of a Bratteli diagram \cite{Bratteli}, which specifies an approximately finite-dimensional (AF) $C^*$-algebra with a representation on a Hilbert space, which is isomorphic to the anyonic fusion Hilbert space.
An AF $C^*$-algebra $\mathcal{A}$ is given by a direct limit $\mathcal{A}=\underrightarrow{lim}\mathcal{A}_{n}$ of a finite dimensional $C^*$-algebra $\mathcal{A}_n$ where $\mathcal{A}_n$ is a direct sum of matrix algebras over $\mathbb{C}$, $\mathcal{A}_{n}=\oplus_{k=1}^{N_{n}}\mathcal{M}_{r_{k}}(\mathbb{C})$. Similarly, a Hilbert space representation of $\mathcal{A}$, $H^{\mathcal{A}}$, is obtained as a direct limit of a system of finite dimensional Hilbert spaces $H_{n}^{\mathcal{A}}$, $H_{n}^{\mathcal{A}}=\oplus_{k=1}^{N_{n}}\mathbb{C}^{r_{k}}$.  A Bratteli diagram yields a unique $C^*$-algebra and allows for a simpler computation of the dimension of the Hilbert space representations of this algebra by counting the number of paths to a certain node. For the Fibonacci $C^*$-algebra, see \cite[Example III.2.6]{Davidson} and \cite[Section 3.2]{Marcolli2015}, for the Bratteli diagram illustration and the dimension of the Hilbert space computation. The isomorphism between the representations of Hilbert spaces and the anyonic fusion Hilbert spaces is given in \cite[Lemma 3.3]{Marcolli2015}, where the dimensions of Fibonacci anyons and the Fibonacci $C^*$-algebra both grow with the Fibonacci sequence.

\subsection{Fibonacci Quasicrystals and the Fibonacci $C^*$-algebra}
\label{subsec:12}

In analogy with the anyonic case, we will provide a physical description of the anyonic behavior of quasicrystals to allow for a concrete physical implementation and then the associated effective fusion Hilbert space to deal with topological quantum information processing. It is more common to deal with quasicrystals from the point of view of Bloch theory for periodic many body atomic quantum systems, but even within this point of view there are different implementations. There is vast literature on computation of the spectrum, band structure and topological properties; we will highlight \cite{Bellissard1992,Kohmoto1986,Sutherland1986,Fujiwara1989,Luck1989,suto1989,benza1991,kaliteevski2000,Florescu2009,Kalugin2014,Tanese2014,Gambaudo2014,Mace2017,Mace2019,Sen2020,Baggioli2020,Jagannathan2021,Indubala2013,Ghadimi2017,Varjas2019,Cao2020,Duncan2020,Liu2021,Hua2021,Fraxanet2021,Rosa2021,Sarangi2021,Fan2022}. From our understanding, the different approaches have convergent results, including the self-similar structure of the energy spectrum, band structure and topological protected phases. The geometric cut-and-project method, or its more general form called model sets, describes this structure. 
The starting point is the periodic Bloch theory considering the Schrodinger equation for a particle over the atomic structure with a periodic potential $V(r+R)=V(r)$ for all lattice vectors
$R$ of a given lattice $\mathcal{L}$. With this setup, the Hamiltonian commutes with the translation operators and the Bloch theory diagonalizes both simultaneously. For this, one introduces the reciprocal lattice $\mathcal{L}^*$ with primitive translation vectors $K$ where the scalar product $R.K$ is an integer multiple of $2 \pi$. The eigenfunctions 
are such that $k$ exists as 
\begin{equation}
\psi_{k+K}(r+R)=e^{ik.R}\psi_{k}(r),
\end{equation}
which $\psi_{k}(r)$ the Bloch wavefunctions on $\mathbb{R}^{n}\times\mathbb{R}^{n}$
($r$ in the Voronoi cell $V$ and $k$ in its dual $V^{*}$, also called Brillouin zone). 
The curves of the spectrum are periodic in dual reciprocal space, and the entire band structure is defined by the band structure inside the first Brillouin zone.

Our idea is to study the Hilbert space of $\psi$'s satisfying 
Bloch's theorem such that $||\psi||^{2}<\infty$. We then introduce
for each $k\in V^*$ the Hilbert space $H_{k}$ of functions $u$ on
$\mathbb{R}^{n}$ such that 
\begin{equation}
u(r+R)=e^{ik.R}u(r),\label{eq.blochwavefunctionk}
\end{equation}
and $||u||^{2}<\infty$, with $H^{\mathcal{L}}=\oplus H_{k}$ and the dimension growing with the number of points on the lattice. The Hilbert spaces for a particle over an aperiodic potential of quasicrystals will be seen as subspaces of these lattice Hilbert spaces $H^{\mathcal{L}}$ and we will need to review the cut-and-project method to get the quasicrystals from the lattice $\mathcal{L}$.

We consider a cut-and-project scheme (CPS) as a 3-tuplet $\mathcal{G}=\left(\mathbb{R}^{d},\mathbb{R}^{d'},\mathcal{L}\right)$,
where $\mathbb{R}^{d}$, called the parallel space, and $\mathbb{R}^{d'}$,
the perpendicular space, are real euclidean spaces and $\mathcal{L}$
is the lattice in $\mathcal{E}=\mathbb{R}^{d}\times\mathbb{R}^{d'}$,
the embedding space, with the two natural projections $\pi$:$\mathbb{R}^{d}\times\mathbb{R}^{d'}\rightarrow\mathbb{R}^{d}$
and $\pi_{\bot}$:$\mathbb{R}^{d}\times\mathbb{R}^{d'}\rightarrow\mathbb{R}^{d'}$,
subject to the conditions that $\pi(\mathcal{L})$ is injective and
that $\pi_{\bot}(\mathcal{L})$ is dense in $\mathbb{R}^{d'}$. With
$L=\pi(\mathcal{L})$, this scheme has a well-defined map called the
star map $\star:L\rightarrow\mathbb{R}^{d'}:$
\begin{equation}
x\longmapsto x^{\star}\coloneqq\pi_{\bot}(\pi^{-1}(x)).\label{eq:starmap}
\end{equation}
For a given CPS $\mathcal{G}$ and a window $W$, quasicrystal point sets ($\triangle_{\gamma}^{\lambda}(W)$) can be generated by setting two additional parameters: a shift 
$\gamma\in\mathbb{R}^{d}\times\mathbb{R}^{d'}/\mathcal{L}$ with $\gamma_{\bot}=\pi_{\bot}(\gamma)$, and a scale parameter $\lambda\in\mathbb{R}$. The projected set
\begin{equation}
\triangle_{\gamma}^{\lambda}(W)\coloneqq\left\{ x\in L\mid x^{\star}\in\lambda W+\gamma_{\bot}\right\} =\left\{ \pi(y)\mid y\in\mathcal{L},\pi_{\bot}(y)\in\lambda W+\gamma_{\bot}\right\} ,\label{eq:modelset}
\end{equation}
gives the quasicrystal point set. 

Another important concept is the one of a tiling of the Euclidean space from the point set. 
Consider that a pattern $\mathcal{T}$ in $\mathbb{R}^{d}$ ($\mathcal{T}\sqsubset\mathbb{R}^{d}$)
is a non-empty set of non-empty subsets of $\mathbb{R}^{d}$. The
elements of $\mathcal{T}$ are the fragments of the pattern $\mathcal{T}$.
A tiling in $\mathbb{R}^{d}$ is a pattern $\mathcal{T}=\{T_{i}\mid i\in I\}\sqsubset\mathbb{R}^{d},$
where $I$ is a countable index set, the fragments $T_{i}$ of $\mathcal{T}$
are non-empty closed sets in $\mathbb{R}^{d}$ subject to the conditions
\begin{enumerate}
\item $\cup_{i\in I}T_{i}=\mathbb{R}^{d}$, 
\item $int(T_{i})\cap int(T_{j})=\textrm{Ø}$ for all $i\neq j$ and 
\item $T_{i}$ being compact and equal to the closure of its interior $T_{i}=\overline{int(T_{i})}$. 
\end{enumerate}
While this is trivial for lattices with a unique unit cell, quasicrystals instead have more than one unit cell. Multiple quasicrystals with the same number of points $N$ from $\mathcal{L}$ projected to the parallel space can lead to different tilings depending on the shift parameter $\gamma$.

The construction above identifies the quasicrystal point set as a subset of the original lattice in the embedding space and its Hilbert space $H^{\triangle}$ as a subspace of the lattice Hilbert space $H^{\mathcal{L}}$. An explicit example is given in \cite[Section 3.2]{Bellissard1992} for the one dimensional Fibonacci chain derived from the $\mathbb{Z}^{2}$ lattice. This provides access to physical properties of quasicrystals, such as the electronic structure. However, the full tiling structure is not properly captured by these descriptions. To address the different tiling configurations of quasicrystals, it is standard to consider the associated $C^*$-algebra structures \cite[Sections II.3 and V.10]{Connes} and the notion of tiling spaces \cite{sadun2018}. A simple way to look at this is to decompose the quasicrystalline Hilbert space $H^{\triangle}$ further according to tile configurations.
The one-dimensional Fibonacci chain and the two-dimensional Penrose tiling can be described by only two tiles. For the Fibonacci chain, they are called long (L) and short (S) edges. For the Penrose tiling, they can be given by two different rhombuses, the thin rhombus (T) and the fat rhombus (F) or two quadrilaterals called kites and darts. 

We can then consider the Hilbert spaces $H^{\triangle}_{L,F}$ and $H^{\triangle}_{S,T}$ associated to the two different tiles. The frequency of appearance of these tiles in some tiling is constant and grows with the Fibonacci sequence, given, at some step, $F(N)$ for L or F to $F(N-1)$ for S or T.  From the Bloch theory, the number of states depends on the number of points in the lattice, which translates to the number of tiles. A lattice trivially has only one tile. For quasicrystals, the number grows differently depending on the tiling considered. Both the Fibonacci chain and Penrose tiling contain two fundamental tiles that grow with the Fibonacci sequence. As such, the Hilbert spaces $H^{\triangle}_{L,F}$ and $H^{\triangle}_{S,T}$ subspaces of a quasicrystalline Hilbert space (which are subspaces of lattices Hilbert spaces), have dimensions growing with the number of tiles added to the quasicrystal in the same way that the dimensions of the anyonic fusion Hilbert spaces grow with the addition of anyons. Following the discussion from the previous section, we conclude that these quasicrystalline subspaces are candidates to implement representations of the Fibonacci $C^*$-algebra associated to Fibonacci anyons. We see the tiles emerging from the Bloch theory playing the same role of the non-abelian $SO(3)$ sources in the Chern-Simons theory.

Another way to look at this is to consider the tiling space, which leads to Hilbert spaces isomorphic of the ones considered above with dimensions growing with the Fibonacci sequence. 
Basically, one starts with a quasicrystal point set $\triangle_{\gamma}$ and associates a tiling with it. Then, we can shift the point set by shifting the window in perpendicular space using $\gamma_{\bot}$. Each shift generates a new tiling with the same tiles but with a different configuration, where this tile can be seen in both parallel and perpendicular spaces due to the star map. The difference is that in parallel space there is a growth of the quasicrystal with tiles of fixed length, while in the perpendicular space each point added rescales the tiles and reorganizes the configuration leading to a rescaling of the space, which is usually called \textit{inflation} or \textit{deflation} for the inverse process. Each tiling is a point in the so-called tiling space, which encodes all possible tilings that can be made with a fixed CPS and window. To encode this information, we can fix a point $x$ inside the window in the perpendicular space. As points are projected, with $\pi_{\bot}(\mathcal{L})$,  
we can track the tile type around $x$ after a new point is projected. Then, we can generate different tilings from different shifts and track the sequence of tiles around that point $x$ over the different sequence of projections. 

\begin{figure}[!h]
	\centering{}
	\includegraphics[scale=0.16]{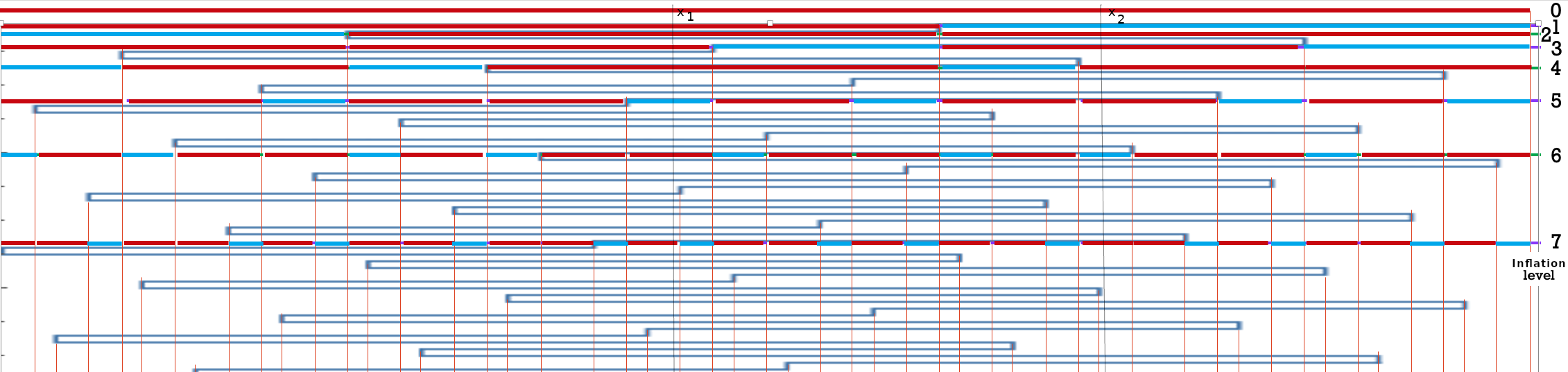}
	\caption{The segment of the window in perpendicular space for the Fibonacci chain is shown at each inflation/deflation level. The $L$ tiles are in red and $S$ tiles in blue. On the horizontal axis, we show specific Fibonacci-chain configurations, where the number of tiles grows with the Fibonacci sequence. The sequences $(x_i)_n$ are given by vertical lines. For example, we show two possible sequences at $x_1$ and $x_2$.}
	\label{perpspaceinflationsxpositions}
\end{figure}

\begin{figure}[!h]
	\centering{}
	\includegraphics[scale=0.16]{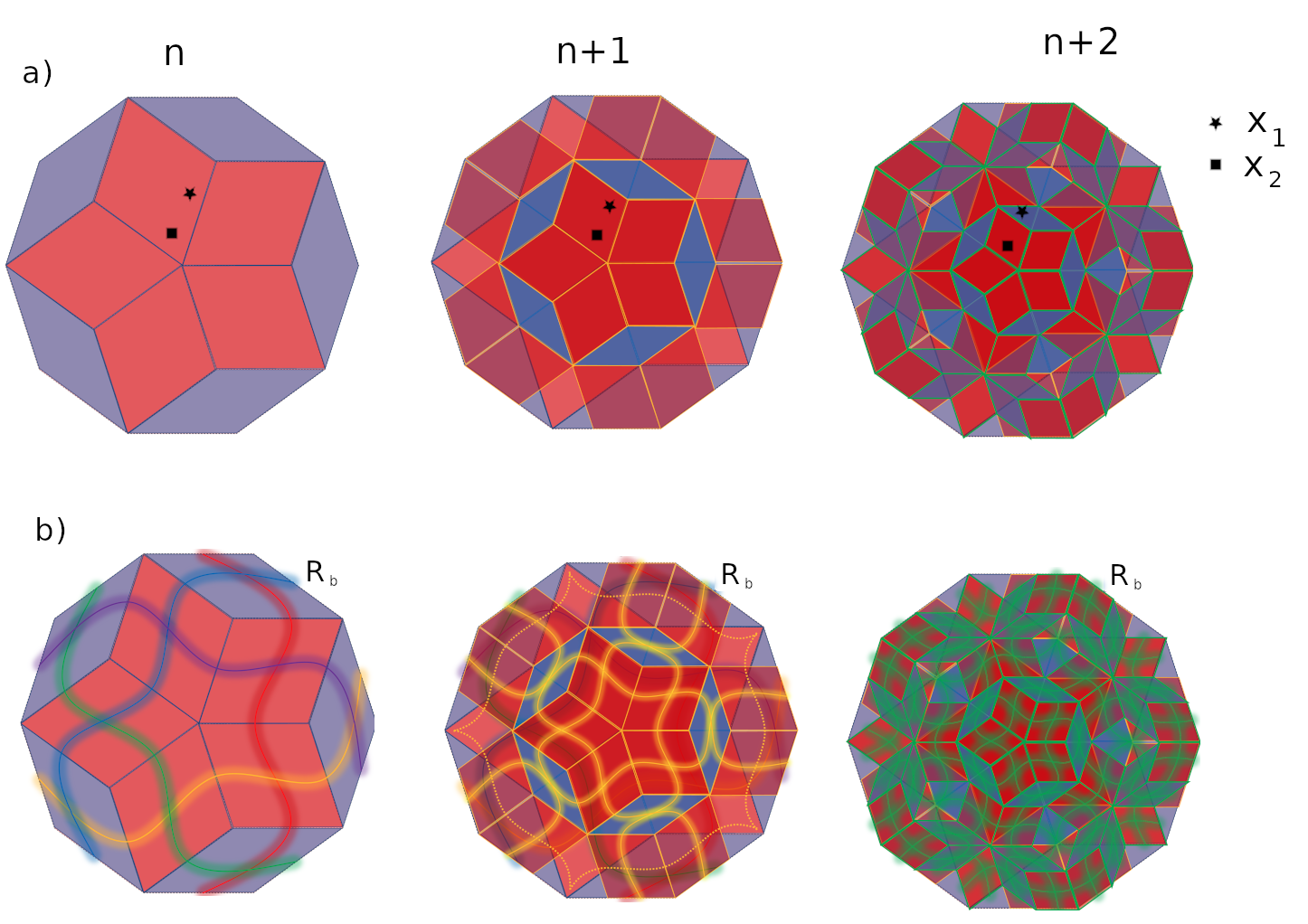}
	\caption{In (a), we show three inflations tracking two positions $x_1=110$, $x_2=111$ over the inflation levels with the fat rhombus in red and the thin in blue. In (b), we introduce the ribbon description. The ribbons are constructed by straight lines (smooth for illustration purposes on the image) going from the center of one tile to the center of an adjacent tile following the Fibonacci rules on the same level as the inflations. For example, the ribbon $R_b$ (the blue in the $n$th level) goes over the following tiles in the three levels shown: $TFFT$, $FTFFTF$ and $FTFTFFTFTF$. Note that a ribbon going over an F in one level will go over an F and T in the next inflation level and a ribbon going over an S will always go to an F.}
	\label{PTinflationdeflationRibbons}
\end{figure}

Equivalently, one can use only one projection and track the evolution for different positions inside the window. Each tiling is described by a sequence that encodes the evolution of tiles around $x$ in the perpendicular space as the quasicrystals grow in parallel space. By labelling the Fibonacci-chain and Penrose-tiling letters L or F as the number 1 and S or T as 0, we can associate different sequences $(x_i)_n$ of 0's and 1's to $x$, where $i$ indexes the different sequences of projections and $n\in\mathbb{N}$ is the level in one sequence of projections. The only constraint on these sequences, which arises from the geometry of the CPS with fixed window, is that if $(x_i)_n=0$, then $(x_i)_{n+1}=1$. 
We illustrate this for the Fibonacci chain in Figure \ref{perpspaceinflationsxpositions}, where $x_1=11111011...$ and $x_2=10111101...$, for example.
And for the Penrose tiling in Figure~\ref{PTinflationdeflationRibbons}, where $x_1=110...$ and $x_2=111...$.
Additionally, an equivalence relation is defined on this space of sequences. Tilings $T_i$ and $T_j$, where there is some $m$, such that $(x_i)_n = (x_j)_n$ for $n \geq m$ are equivalent. This is presented in detail in \cite[Sections II.3 and V.10]{Connes} for the tiling space of the Penrose tiling with the construction of a $C^*$-algebra $\mathcal{A}$ associated to this space. Remarkably, this algebra is the same Fibonacci $C^*$-algebra; the Hilbert space representations are isomorphic to the anyonic fusion Hilbert spaces \cite{Marcolli2015}. 
In the next section, we present detailed aspects of this algebra, quasicrystal physics interpretations, and topological quantum computation.

\begin{figure}[!h]
	\centering{}
	\includegraphics[scale=0.22]{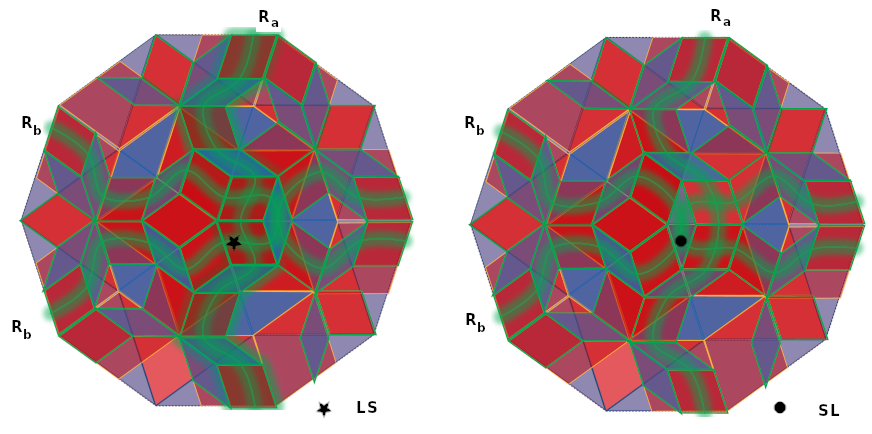}
	\caption{Tile flip that send ribbons $R_b$ from FTFTF\textbf{FT}FTF to FTFTF\textbf{TF}FTF given a factor of $\phi^{-2}$ on the associated states. The Ribbon $R_a$ has a change in orientation on the flip position.}
	\label{ribbonphasonflip}
\end{figure}

Let us consider a concrete solution of a Hamiltonian for a quasicrystal. Despite the difficulties with the generalization of the Bloch and Floquet's theories, there are a few known exact solutions for quasicrystal Hamiltonians. Some of the state solutions of the so-called tight-binding model for the Fibonacci chain and the Penrose tiling are known \cite{Kohmoto1986,Sutherland1986,Fujiwara1989,Kalugin2014,Mace2017,Jagannathan2021}. These states include zero-energy degenerate states and have a similar form of the Bloch wave function, Eq. (\ref{eq.blochwavefunctionk}), given by 
\begin{equation}
\psi(i)=C(i)e^{\kappa h(i)}\label{eq:exactstateqc}
\end{equation}
where $\kappa\in\mathbb{R}$ is a constant, $C(i)$ are local site-dependent periodic functions given the local amplitudes and $h(i)$ is a non-local field, so-called height field, depending on the geometry of the specific tiling. For the Fibonacci chain in Eq. (\ref{eq:exactstateqc}), the zero energy state takes the form $\psi(2i)=(-1)^{i}e^{\kappa h(2i)}$ with $\kappa=\ln \phi$ and the field $h(2i)$ given by
\begin{equation}
h(i)=\sum_{0\leq j\leq i}B(2j\rightarrow2(j+1)),
\end{equation}
with $B(LS)=1$, $B(SL)=-1$ and $B(LL)=0$. For the Penrose tiling, both $\kappa$ and $C(i)$ are computed numerically \cite{Mace2017} but the ribbon description discussed above allow us to access the Fibonacci chain subspaces directly. Note that a flip $LS\rightarrow SL$, such as the the one for the ribbon $R_b$ in Figure~\ref{ribbonphasonflip}, changes the state by a factor of $\phi^{-2}$, $\psi^{LS}(i)=\phi^{-2}\psi^{SL}(i)$.

\section{Quasicrystalline Topological Quantum Information Processing}
\label{sec:2}

Following the Bloch theory, a quantum-mechanical quasicrystal is described by a Hilbert space, which is a subspace of a Hilbert space describing a higher-dimensional crystal (the lattice $\mathcal{L}$ from the previous section). 
In principle, this gives us a mechanism to grow a quasicrystal maintaining the quantum superposition of tilings in a tiling space. This growth is described by the sequences of 0's and 1's (encoding the different two tiles in the Fibonacci chain or Penrose tiling) $(x_i)_n$, such that if $(x_i)_n=0$, then $(x_i)_{n+1}=1$ and subject to some equivalence relation, such as the one described in the previous section with one associated algebra $\mathcal{A}$. A slightly different, but equivalent way to address the tiling space is to consider finite sequences $(x_i)_n$, $n=1,...,N$, subject to the same rule and with a equivalence relation given by $(x_i)_N = (x_j)_N$ and construct the algebra $\mathcal{A}$ as the inductive limit of finite-dimensional algebras $\mathcal{A}_N$ with $\mathcal{A}_N$ as a direct sum of matrix algebras \cite{Tasnadi2002}. For the Fibonacci chain and Penrose tiling described by just two tiles, the set of equivalence classes has only two elements, with the number of both tiles growing with the Fibonacci sequence (for example $L$ grows with $F(N+1)$ and $S$ with $F(N)$), which gives $\mathcal{A}_{N}=M_{d_L^n}\oplus M_{d_S^n}$ with $d_l^n=F(N+1)$ and $d_S^n=F(N)$.  
The embedding of $\mathcal{A}_{N}$ in $\mathcal{A}_{N+1}$ given by $d^{n+1}_L=d_L^n+d_S^n$ and $d_S^{n+1} = d_L^n$.
To do the inverse process and merge tiles, one can define a projection at the step $N$, by means of the operation to forget that step, remaining with sequences with $n=1,...,N-1$.

One can then consider projections $E_n$, acting on the associated Hilbert spaces defined by the $\mathcal{A}_{N}$ such that $E_n$ maps the Hilbert space $H_{d_L^n}$ to $H_{d_L^{n-1}}$ or subspaces of $H_{d_L^n}$ associated to $A_N$ to subspaces of $H_{d_L^{n-1}}$ associated to $A_{N-1}$ \cite{Jones1983}.
Following \cite[Lemma 5 in section V.10]{Connes}, we consider a sequence of $E_n$ orthogonal projections, known as Jones-Wenzl projections, such that the following relations hold
\begin{eqnarray}
E_{n}^{2} &=& E_{n} \label{eq:quasicrystalJonesAlgebra0} \\ 
E_{n}E_{m}E_{n} &=& \phi^{-2}E_{n} \quad \mbox{if }\thinspace|n-m|=1 \label{eq:quasicrystalJonesAlgebra1} \\ 
E_{n}E_{m} &=& E_{m}E_{n}  \quad \mbox{ if }\thinspace|n-m|>1,\label{eq:quasicrystalJonesAlgebra}
\end{eqnarray}
where for more general quasicrystals, one could consider Eq. (\ref{eq:quasicrystalJonesAlgebra1}) as $E_{n}E_{m}E_{n} = [2]_q^{-2}E_{n}$ with the so-called quantum numbers $[n]_q$ given by
\begin{equation}
\left[n\right]_{q}=\frac{q^{n}-q^{-n}}{q-q^{-1}}\label{eq:qnumber}
\end{equation}
with $q=e^{\frac{\pi i}{r}}$. In the case of Eqs. (\ref{eq:quasicrystalJonesAlgebra0}-\ref{eq:quasicrystalJonesAlgebra}), we have $q$ a fifth root of unity, $r=5$, and we call the algebra $\mathcal{A}_N(q)$.

In the study of Fibonacci anyons, the Temperley-Lieb algebra with generators $F_n$ is typically used such that $E_n=\phi^{-1}F_n$, see \cite[Section 8.2.2]{Pachos} and \cite{Kauffman2018}. The algebra defined by the projections $E_n$, Eqs. (\ref{eq:quasicrystalJonesAlgebra0}-\ref{eq:quasicrystalJonesAlgebra}), is isomorphic to the Fibonacci $C^*$-algebra of Fibonacci anyons and Fibonacci quasicrystals, the proof can be seen by explicitly deriving its Bratteli diagram \cite{Jones1983}.
The quasicrystal projections can be used to implement the braid operations necessary for quantum evolution to implement topological quantum computing. In the case of anyons, moving one anyon around the other is a non-trivial operation encoded in the braid group operations on the fusion Hilbert space. For non-abelian anyons, these operations are shown to be dense in $SU(N)$, with $N$ as the number of anyons on the system to provide universal quantum computation. The braid group is generated by generators $B_n$ satisfying the relations
\begin{eqnarray}
B_{n}B_{n}^{-1} &=& B_{n}^{-1}B_{n}\nonumber \\
B_{n}B_{m}B_{n} &=& B_{m}B_{n}B_{m} \quad \mbox{if } \thinspace|n-m|=1\nonumber \\
B_{n}B_{m} &=& B_{m}B_{n} \qquad \mbox{ if } \thinspace|n-m|>1.\label{eq:braidgroup}
\end{eqnarray}

A representation of the braid group can be given from the algebra in Eq. (\ref{eq:quasicrystalJonesAlgebra}) by 
\begin{align}
\rho_{A}(B_{n}) & =\phi AE_{n}+A^{-1}\mathbb{I}\nonumber \\
\rho_{A}(B_{n}^{-1}) & =\phi A^{-1}E_{n}+A\mathbb{I},\label{eq:qcbraid}
\end{align}
with $\phi=-A^2-A^{-2}$, where unitarity is guaranteed if the projections $E_n$ are Hermitian. $A$ contains four solutions, all with $|A|=1$. The four solutions are $A = e^{3\pi i/5}$, $-e^{3\pi i/5}$, $e^{2\pi i/5}$, and $-e^{2\pi i/5}$. Note that the $R$-matrix for Fibonacci anyons in Eq.~\eqref{eq:RFmatrices} contains $e^{3\pi i/5}$ on some of the diagonals. With the solution of $A$ provided, one can verify that
\begin{eqnarray}
\rho_A(B_{n}) \rho_A(B_{n}^{-1}) &=& \rho_A(B_{n}^{-1}) \rho_A(B_{n})\nonumber \\
\rho_A(B_{n})\rho_A(B_{m})\rho_A(B_{n}) &=& \rho_A(B_{m}) \rho_A(B_{n}) \rho_A(B_{m}) \quad \mbox{if } \thinspace|n-m|=1\nonumber \\
\rho_A(B_{n}) \rho_A(B_{m}) &=& \rho_A(B_{m}) \rho_A(B_{n}) \qquad\qquad \mbox{ if } \thinspace|n-m|>1.
\end{eqnarray}
Therefore, the quasicrystal projection operators can be used to construct a representation of the braid group.

The usual step from quantum computation to topological quantum computation can now be done with quasicrystals by finding an embedding $\mathfrak{e}$ of an $N$-qubit space $(\mathbb{C}^{2})^{\otimes N}$ into a subspace of the tiling space. The embedding does not need to be efficient because it is well known that the braid group can approximate any universal quantum gate to desired precision. The computational subspace of the tiling space can be given by fixing to one equivalence class $(x_i)_n$, $n=1,...,2N+1$ and $i=1,...,d$ with $d$ the number of sequences with $(x_i)_{2N+1} = 1$. We represent this subspace by $T_{N,1}=(x_i)_n$. Finally, to simulate a quantum circuit, we can have
\begin{eqnarray}
\left(\mathbb{C}^{2}\right)^{\otimes N} & \rightarrow^{\mathfrak{e}} & T_{N,1}\nonumber \\
U\downarrow &  & \downarrow\rho_{A}(B)\nonumber \\
\left(\mathbb{C}^{2}\right)^{\otimes N} & \rightarrow^{\mathfrak{e}} & T_{N,1} . \label{qubitanyonmapap}
\end{eqnarray}
Explicit matrix representations of $\rho_{A}(B)$ can be obtained from the algebra $\mathcal{A}_N(q)$ acting on the $N$-qubit Hilbert space $(\mathbb{C}^{2})^{\otimes N}$, a subspace of the tiling space. Define $E(q)$ acting on $\mathbb{C}^{2}\otimes\mathbb{C}^{2}$ as \cite{Goodman1993}
\begin{equation}
E(q)=[2]_{q}^{-1}\left(q^{-1}e_{11}\otimes e_{22}+qe_{22}\otimes e_{11}+e_{12}\otimes e_{21}+e_{21}\otimes e_{12}\right)\label{eq:Eqonqubitspace}
\end{equation}
with $e_{ij}$ the 2 dimensional matrix units and then $E_{i}(q)=\mathbb{I}\otimes...\otimes\mathbb{I}\otimes E(q)\otimes...\otimes\mathbb{I}$, where 
$E(q)$ acts in the positions $i$ and $i+1$ of the tensor places.

\begin{figure}[!h]
	\centering{}
	\includegraphics[scale=0.30]{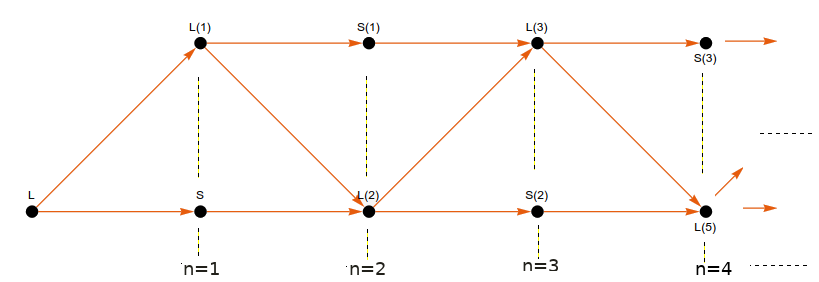}
	\caption{A Bratteli diagram for the Fibonacci chain (similar for the Penrose tiling with fat (F) and thin (T) rhombus), where each path, $i$, to a node gives a $x_i$ and the different inflations levels $n$ is shown. The number in parentheses is the number of paths to that node at level $N$, $n=1,...,N$, which gives the Hilbert space dimension for the associated subspace with sequences $(x_i)_N=$ $L$ or $S$. }
	\label{brattelidiagramfibchain}
\end{figure}

For TQC with a quantum-mechanical quasicrystal, suppose that experimentalists in the future could have complete control of how the quasicrystal is grown. The number of possible inflation/deflation paths in tiling space, which gives the Hilbert space dimension, is tied to the number of physical tiles, analogous to how the number of physical anyons define the fusion Hilbert space dimension. This allows us to get a dictionary between concepts related to Fibonacci anyons and TQC with a quantum-mechanical quasicrystal. For concreteness and simplicity, consider the Fibonacci chain, which has two inflation rules
\begin{eqnarray}
\mbox{Rule A: } && \{\mbox{L}\rightarrow \mbox{LS}, \mbox{ S} \rightarrow \mbox{L} \} \nonumber \\
\mbox{Rule B: } && \{\mbox{L}\rightarrow \mbox{SL}, \mbox{ S} \rightarrow \mbox{L} \}.
\end{eqnarray}
To clarify, our conventions are that the inflation rules apply an inflation. 
It can be verified that successive application of Rule A seeded by S leads to the reverse of the chain found by successive application of Rule B. If $n$ arbitrary combinations of Rule A and Rule B are applied from the seed, then $2^n$ states can be found. However, these lead to various duplicate tilings, such that $Fib(n+2)$ unique tilings are found. For example, with seed $L$, for $n=2$ we have \{\{L,SL,LSL\}, \{L,SL,LLS\}, \{L,LS,SLL\}, \{L,LS,LSL\}\} resulting in 3 unique states \{LSL, LLS, SLL\} or in terms of the $(x_i)$, $i=1,2,3$, describing the associated tiling space, we have \{LSL, LLS, LLL\}. The associated Bratteli diagram is shown in Figure~\ref{brattelidiagramfibchain}, which is equivalent to the Fibonacci anyon diagram \cite{Marcolli2015} and the $\mathcal{A}_N(q)$ diagram for Jones-Wenzl projections \cite{Jones1983}.

The analogue of an anyonic fusion process is given by the operation to forget the $N$th step in $(x_i)_n$, $n=1,...,N$, leaving the sequences $(x_i)_{n}$ with $n=1,...,N-1$. This sends the system from level $N$ to $N-1$ or the Hilbert space of dimension $F(n)$ to $F(n-1)$ and is equivalent to a deflation in the physical quasicrystal. Since L is a fixed length, this operation acting on the Hilbert space associated with the two tiles LS would lead to L as an deflation, which decreases the length of the chain.
When performing the analogue of braiding in the quasicrystal, one specifies a basis given by inflation/deflation paths $(x_i)_{n}$ and decomposes the projection $E_n$ in a direct sum of projections acting in lower dimensional subspaces. From Eq.~\eqref{eq:qcbraid}, the subspace acted by $E_n$ gets a different phase, which relates to $A$, and a rescaling by $\phi$. In usual anyonic systems, the braid operations involve a basis transformation. This selects two anyons to be fused and applies an operation on these two anyons, which gives a phase $R$, and then applies an inverse basis transformation. In quasicrystals, the projection $E_n$ selects directly the subspace to be acted by a phase and rescaling.
Table \eqref{tab:Dictionary} summarizes a dictionary that compares aspects of Fibonacci anyons and quantum-mechanical Fibonacci chains for TQC. 
\begin{table}
    \centering
    \begin{tabular}{c|c}
        Fibonacci Anyons & Quantum Fibonacci Chain \\ \hline
        Anyon & Tile \\
        0, 1 & S, L \\
        $d$-fold degeneracy & \# of tiles \\ 
        Fusion with 1 (anyon destruction) & Deflation (tiles merging) \\ 
        Braid $B=FRF^{-1}$ & $\rho_{A}(B_{n})= A \phi E_{n}+A^{-1}\mathbb{I}$ 
    \end{tabular}
    \caption{A dictionary between concepts related to Fibonacci anyons and TQC with a quantum-mechanical Fibonacci chain is provided.}
    \label{tab:Dictionary}
\end{table}

We have already noted that crystallographic theories, mainly Bloch and Floquet's theory, do not extend directly to quasicrystals due to the lack of translational symmetry. We also discussed an isomorphism between anyonic and quasicrystalline Hilbert spaces. In this context, it is tempting to import well-developed techniques from anyonic systems for applications in quasicrystals to implement TQC. One example is the so-called golden chain \cite{Feiguin2007} modeling Fibonacci anyons in one dimension. The golden chain has a natural realization in terms of the Fibonacci-chain quasicrystal. The local Hamiltonian $H_i$ acting on the $i$th Fibonacci anyon on the chain discussed in \cite{Feiguin2007} is immediately identified with the projections $E_n$, acting on the inflation level $n$, $(x)_n$, of the Fibonacci-chain quasicrystal, allowing access to the quantum quasicrystal growth and shrinkage.  A detailed analysis of this Hamiltonian (and other anyonic Hamiltonians) in context of quasicrystals and their relationship with quasicrystal Hamiltonians could be discussed in future work.

\section{Implications}
\label{disc}

Conceptually, topological quantum computing is known to have advantages over standard quantum computing for scaling due to hardware-level error protection. But the physical implementation of topological phases of matter is a big challenge. One main line of research is to implement localized Majorana modes, which behave as abelian Ising anyons. This line of research has seen a major setback recently, with a main group of researchers withdrawing papers that claimed experimental validation of abelian anyons, in particular the Majorana fermions excitations \cite{Zhang2021,Gazibegovic2017}. This opens the opportunity for new approaches to topological quantum computing through the discovery of new hardware platforms that can support the anyonic quantum information processing. 

The novelty of our work is the proposal of quasicrystal materials as such a natural platform. These materials exhibit aperiodic and topological order and they are already implemented in laboratories around the world. More difficult is the manipulation of topological properties of tiling spaces of quasicrystals required to the task of quantum information processing, to which our work adds further theoretical understanding.

\end{document}